\documentclass[groupedaddress, beta
3),10pt,twocolumn,bibnotes,prl,showpacs]{revtex4}%
\usepackage{amssymb}%
\usepackage{amsmath}%
\usepackage{amsfonts}%
\usepackage{graphicx}

\begin{document}
\title{Unifying Entanglement and Nonlocality as a Single Concept: Quantum Wholeness}
\author{Zeng-Bing Chen}
\author{Sixia Yu}
\affiliation{Department of Modern Physics, University of Science and Technology of China,
Hefei, Anhui 230026, People's Republic of China}
\author{Yong-De Zhang}
\affiliation{Department of Modern Physics, University of Science and Technology of China,
Hefei, Anhui 230026, People's Republic of China}
\author{Nai-Le Liu}
\affiliation{Department of Modern Physics, University of Science and Technology of China,
Hefei, Anhui 230026, People's Republic of China}
\thanks{To whom correspondence should be addressed. E-mail: zbchen@ustc.edu.cn}

\begin{abstract}
Although entanglement is widely recognized as one of the most fascinating
characteristics of quantum mechanics, nonlocality remains to be a big
labyrinth. The proof of existence of nonlocality is as yet not much convincing
because of its strong reliance on Bell's theorem where the assumption of
realism weakens the proof. We demonstrate that entanglement and quantum
nonlocality are two equivalent aspects of the same quantum wholeness for
spacelike separated quantum systems. This result implies that quantum
mechanics is indeed a nonlocal theory and lays foundation of understanding
quantum nonlocality beyond Bell's theorem.

\end{abstract}
\pacs{03.65.Ud, 03.67.-a, 03.65.Ta}
\pacs{03.65.Ud, 03.67.-a, 03.65.Ta}
\maketitle

Since the classic works of EPR (Einstein, Podolsky and Rosen \cite{EPR}) and
Bell \cite{Bell,Bell-book}, quantum nonlocality and entanglement have been
recognized as two crucial notions in modern understanding of quantum
phenomena. As entanglement is an essential \textquotedblleft
resource\textquotedblright\ in practical applications of quantum information
\cite{QIT}, various aspects \cite{QIT,Vedral,Horodecki-e,PhysToday}\ of it
have been extensively studied in recent years. However, our current
understanding \cite{Barrett}\ of nonlocality is heavily based on Bell's
theorem \cite{CS-exp,Peres-book,Werner-rev,Laloe}. Since Werner's seminal work
\cite{Werner}, many elegant ideas
\cite{Popescu,Gisin-hidden,Zuk-Horo,Peres-coll,Werner-rev,Horodecki-e,Barrett-POVM}%
\ have been proposed to understand nonlocality of mixed states. Yet, the
relationship between nonlocality and entanglement for mixed states is still
unclear \cite{Horodecki-e,Werner-rev,Barrett-POVM}\ and remains one of the
most challenging problems in the field. Here we first give a generic locality
condition (i.e., a condition that should be satisfied by any local theory and
can thus be regarded as a proper definition of locality) for any multipartite
system (with its subsystems being spacelike separated). We then prove that the
locality condition, also underlying Bell's inequalities, is satisfied if and
only if the states of any spacelike separated quantum system are
non-entangled. As such, any entangled state is quantum mechanically nonlocal,
i.e., entanglement and quantum nonlocality are two side of the same attribute
of entangled systems. It is anticipated that the result may be a starting
point for penetrating the mystery of nonlocality of nature in general and
quantum nonlocality in particular.

The concept of entanglement is well justified in the sense that one, at least,
has a definition of entangled states \cite{Werner}. Quantification
\cite{Vedral} and manipulation \cite{PhysToday} of entanglement are hot topics
of fundamental interest in the fields of quantum mechanics and quantum
information theory; various separability criteria \cite{Horodecki-e} have been
found to classify quantum states into entangled and separable ones. By
contrast, quantum nonlocality still remains, to a large extend, an intuitive
notion. Roughly spoken, it usually means the impossibility of simulating
certain quantum predictions by local realistic theories \cite{Barrett}. As is
well known, local realism represents a world view \cite{CS-exp} which states
that a physical system has local objective properties, independent of any
observations on other spacelike separated systems. By assuming locality and
realism, the celebrated Bell inequalities (as well as their various
generalizations \cite{CS-exp,Werner-rev,Gisin-Peres,Collins,Chen}) can be
derived and impose an upper bound on correlations of the results of the
\textquotedblleft Bell experiments\textquotedblright. Astonishingly, the upper
bound is violated quantum mechanically by a large class of entangled states
\cite{CS-exp,Werner-rev,Gisin-Peres,Collins,Chen}, as confirmed by many
experiments \cite{CS-exp,Aspect,Pan-GHZ}. To Bell \cite{Bell-book}, the
quantum violations of Bell's inequalities imply that quantum mechanics is a
nonlocal theory. This attitude is largely accepted in literature and serves as
the basis of our current understanding of nonlocality.

However, the attitude is not unquestionable. Anyway, Bell's inequalities are
derived from two underlying assumptions: locality and realism. Thus, the
experimentally confirmed conflict between quantum mechanics and local realism
could be equally interpreted as implying that quantum mechanics is a local
non-realistic theory. On the one hand, it is natural for some authors (see,
e.g., Ref. \cite{Braunstein}) to refute realism, rather than locality, in
quantum mechanics. On the other hand, there are also attempts \cite{Stapp} to
prove the nonlocality of quantum mechanics without explicitly assuming
realism. Yet, the proof is based on some counterfactual reasonings that are
controversial (see Ref. \cite{Vaidman} and references therein). Currently,
whether quantum mechanics is indeed a nonlocal theory remains an open
question, whose answer can be approached only when we have a solid foundation
of nonlocality.

Now it is a well established fact that biparticle
\cite{Gisin-Peres,Collins,Chen} and multiparticle \cite{Popescu-PLA} entangled
pure states always lead to certain violation of Bell-like inequalities. This
convincingly indicates that there may exist a close and direct relationship
between entanglement and nonlocality. Yet, for mixed states the situation is
very puzzling
\cite{Werner,Popescu,Gisin-hidden,Zuk-Horo,Peres-coll,Werner-rev,Horodecki-e,Barrett-POVM}%
. Werner \cite{Werner} first demonstrated that there are entangled states (the
Werner states) that do not violate any Bell inequality. Subsequent works show
that the Werner states possess \textquotedblleft hidden
nonlocality\textquotedblright\ \cite{Popescu,Gisin-hidden}, which, however,
can be uncovered only by invoking generalized measurements
\cite{Popescu,Zuk-Horo,Peres-coll,Barrett-POVM} instead of the standard von
Neumann measurement. Despite of many elegant efforts
\cite{Popescu,Gisin-hidden,Zuk-Horo,Peres-coll,Werner-rev,Horodecki-e,Barrett-POVM}
along the line of the \textquotedblleft Bell paradigm\textquotedblright, there
are still some open problems \cite{Horodecki-e,Werner-rev,Barrett-POVM} on the
relationship between nonlocality and entanglement. The difficulty encountered
in our current understanding of nonlocality might suggest that Bell's theorem
is not sufficient for uncovering nonlocality of quantum mechanics. Then a
fundamental problem arises here: In what sense we classify quantum states into
local and nonlocal ones?

In this work we provide answers to the above-mentioned open questions. This is
achieved by understanding quantum nonlocality at a deeper level going beyong
Bell's theorem. As one might intuitively envision, nonlocality inherent in
entanglement is a purely quantum phenomenon. Meanwhile, realism underlying
Bell's theorem is a world view that logically has nothing to do with quantum
mechanics. Thus, in order to reveal nonlocality as a fundamental feature of
nature, it is necessary to discard the premise of realism; what concerns us
first is the condition that any local theory must satisfy.

Local causality (or simply, locality) means that the experimental results
obtained from a physical system at one location should be independent of any
observations or actions made at any other spacelike separated locations.
Unfortunately, the locality condition
\cite{Bell,Bell-book,Werner-rev,Shimony,assume} was previously given in
company with realism. Since here one needs a locality condition without any
pre-assumption other than locality, the condition must be expressed by
quantities that are experimentally observable for localists.%
\begin{figure}
[ptb]
\begin{center}
\includegraphics[
height=1.91in,
width=2.2328in
]%
{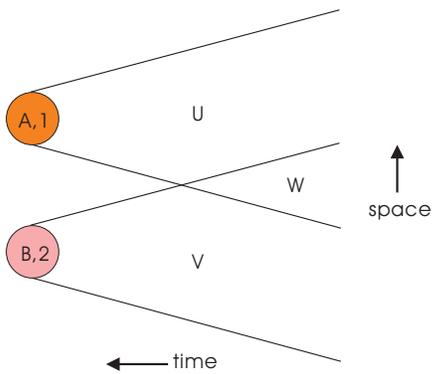}%
\caption{The spacetime diagram of two spacelike separated particles. The
particles A and B are located in two coloured locations which are spacelike
separated. $W$ denotes the overlap of the backward light cores of the
particles A and B; the remaining parts of the backward light cores of the
particles A and B are denoted, respectively, by $U$ and $V$, which are
spacelike separated such that no causal signal can connect them.}%
\label{figure}%
\end{center}
\end{figure}

In order to obtain the desired locality condition in mathematical terms, let
us consider two particles A and B in the spacetime diagram shown in Fig. 1.
They are located, respectively, in two spacetime regions denoted by 1 and 2,
which are spacelike separated. According to local causality, events occurring
in the backward light core of a particle (e.g., particle A or B) may affect
the events (e.g., detecting measurement results) occurring on the particle.
Thus, events in the overlap $W$\ of the backward light cores of the two
particles may be \textquotedblleft common causes\textquotedblright%
\ \cite{Bell-book} of the events in regions 1 and 2, though events in region 1
should not be causes of events in region 2 (and vice versa) as required again
by local causality.

Without loss of generality, one can assume that the observables of A and B
have only a finite number of outcomes. Then consider local measurements on the
two particles. By the standard rule of probability, the requirement of
locality can be mathematically expressed as
\begin{equation}
P(a_{i},b_{j}\left\vert C)\right.  =P(a_{i}\left\vert C)\right.
P(b_{j}\left\vert C)\right. \label{local}%
\end{equation}
for all observed results in the spacelike separated regions 1 and 2 and for
any common causes $C$ in $W$. Here $P(a_{i},b_{j}\left\vert C)\right.
$\ ($i,j=1$, $2$, $3$,\ldots) is a joint probability of measuring any
observables in region 1 (with outcome $a_{i}$) and in region 2 (with outcome
$b_{j}$) conditioned on a given element in the set $C$ of all common causes in
$W$; $P(a_{i}\left\vert C)\right.  $\ [$P(b_{j}\left\vert C)\right.  $] is the
local probability of getting the outcome $a_{i}$ ($b_{j}$)\ when measuring the
observable in region 1 (region 2) conditioned on the given common cause in
$C$. Thus, equation (\ref{local}) has a physically apparent interpretation in
accord with locality: The given common cause can affect the probabilities with
regard to particles A and B; conditioned on the same cause, measurements in
region 1 and region 2 must be mutually independent effects. Importantly, the
common causes in our analysis represent physically a set of certain individual
events whose occurrences can be assigned corresponding probabilities.

Denoting the joint probability of getting outcomes $a_{i}$ and $b_{j}$ as
$P(a_{i},b_{j})$ and the probability of common causes in $W$ as $P(C)$ (see
Fig. $1$), then the rule of conditional probability gives
\begin{equation}
P(a_{i},b_{j})=\sum_{C}P(a_{i},b_{j}\left\vert C)\right.  P(C)\label{pab}%
\end{equation}
where the summation may also mean integration, if necessary. Let us denote the
set of common causes in $W$ by $C=\{\mu\left\vert \mu=1,2,3,\cdots\right.
\}$, with $\lambda_{\mu}\geq0$ being the probability for the cause $\mu$ to
occur and $\sum_{\mu}\lambda_{\mu}=1$. Thus, for a given cause $\mu$, equation
(\ref{local}) becomes $P(a_{i},b_{j}\left\vert \mu)\right.  =P(a_{i}\left\vert
\mu)\right.  P(b_{j}\left\vert \mu)\right.  $.

Obviously, equation (\ref{pab}) can be assigned an operational meaning.
Namely, one can either measure directly $P(a_{i},b_{j})$, or monitor the
common causes first. Conditioned on a specific common cause ($\mu$, say) being
detected, $P(a_{i},b_{j}\left\vert C)\right.  $\ is then measured. Theory of
probability insures that the two procedures are equivalent. Of course, for a
localist, equation (\ref{local}) should be right.

Some further comments in support of the locality condition (\ref{local}) are
noteworthy. First of all, for localists the probabilities in the locality
condition are all measurable quantities and can be measured, in principle,
with an arbitrary accuracy. Given a statistical ensemble, experimenters can
always observe the relative frequencies among all the outcomes; After the
number of the observations tends to infinity, the relative frequencies will
tend to the true probabilities of the corresponding outcomes. Moreover and
more importantly, the locality condition given in a probabilistic terms is
presented without resorting to any specific theory (realistic or quantum) as
it only involves observable quantities; theory enters the picture when one
predicts the probability of each outcome. This feature of the locality
condition enables one to test locality versus any specific theory, i.e., to
test whether nature is local \cite{Chen-test}. For instance, when one uses
hidden-variable theories to assign the probabilities in equation
(\ref{local}), the locality condition (\ref{local}) reduces exactly to Bell's
locality condition \cite{Bell,Bell-book,Werner-rev,Shimony,assume}. Then
following Bell's reasoning yields the usual Bell inequalities which must be
satisfied by any local realistic theory \cite{Bell,Bell-book}. Bell's locality
condition \cite{Bell-book} has been well justified in various aspects in the
context of Bell's inequalities \cite{Werner-rev,Shimony,assume} and is now
widely accepted.

The generic locality condition (\ref{local}) allows one to use quantum
mechanics, instead of hidden-variable theories, when predicting the local
probabilities [e.g., $P(a_{i}\left\vert \mu)\right.  $] appearing in the
locality assumption. As such, the two particles A and B must then be described
locally by standard quantum mechanical rules. Whether the events in $W$\ are
described quantum mechanically or not is unimportant in the subsequent
consideration. When described quantum mechanically, the events in $W$\ may be
represented by quantum states of an \textquotedblleft
ancilla\textquotedblright\ that can be any kind of quantum systems possibly
coupled to both A and B. And in this case, the ancilla will be
\textquotedblleft traced out\textquotedblright\ as one is concerned only with
particles A and B.

Thus in the present context, quantum mechanics can predict these
probabilities, namely, it guarantees that \cite{Peres-book}, for the
probabilities $P(a_{i},b_{j})$, $P(a_{i}\left\vert \mu)\right.  $\ and
$P(b_{j}\left\vert \mu)\right.  $, there are corresponding density operators
$\rho_{AB}$ for the two-particle system, $\rho_{A\mu}$ for A and $\rho_{B\mu}%
$\ for B such that $P(a_{i},b_{j})=Tr(\rho_{AB}\hat{P}_{Ai}\hat{P}_{Bj}%
)=\sum_{\mu}\lambda_{\mu}P(a_{i},b_{j}\left\vert \mu)\right.  $,\ $P(a_{i}%
\left\vert \mu)\right.  =Tr(\rho_{A\mu}\hat{P}_{Ai})$\ and $P(b_{j}\left\vert
\mu)\right.  =Tr(\rho_{B\mu}\hat{P}_{Bj})$. Then equation (\ref{pab})
immediately gives
\begin{equation}
\mathrm{Tr}(\rho_{AB}\hat{P}_{Ai}\hat{P}_{Bj})=\sum_{\mu}\lambda_{\mu
}\mathrm{Tr}(\rho_{A\mu}\rho_{B\mu}\hat{P}_{Ai}\hat{P}_{Bj})\label{roab}%
\end{equation}
Here $\hat{P}_{Ai}=\left\vert a_{i}\right\rangle \left\langle a_{i}\right\vert
$ and $\hat{P}_{Bj}=\left\vert b_{j}\right\rangle \left\langle b_{j}%
\right\vert $\ are the projection operators corresponding to the outcomes
$a_{i}$ and $b_{j}$, respectively. Due to the arbitrariness of $\hat{P}_{Ai}$
and $\hat{P}_{Bj}$, one must have (see also \cite{note})
\begin{equation}
\rho_{AB}=\sum_{\mu}\lambda_{\mu}\rho_{A\mu}\rho_{B\mu}\label{sep}%
\end{equation}
which represents the state of the two particles allowed by the locality
condition (\ref{local}). The state $\rho_{AB}$ is a convex sum, with weights
$\lambda_{\mu}$, of direct products of the local density operators $\rho
_{A\mu}$ and $\rho_{B\mu}$. Different set of common causes leads to different
convex sum of the local density operators.

Two remarks are in order here. The state $\rho_{AB}$ in equation (\ref{sep})
is called \textquotedblleft classically correlated\textquotedblright%
\ \cite{Werner}. Apparently, this classical correlations can only originate
from the `common history' of the particles A and B, as can be seen from the
spacetime diagram (Fig. $1$). In the particular case where the observed
results in $U$ or/and $V$ are independent on $W$, then one always has
$\rho_{AB}=\rho_{A}\rho_{B}$. In this case, $\rho_{A}$ and $\rho_{B}$ contain
already a complete specification of the states of the corresponding particles;
supplementary information from $W$ is redundant, similarly to the classical
case considered by Bell \cite{Bell-book}.

Inverting the above reasoning, it is easy to see that any state given in
equation (\ref{sep}) satisfies locality. Thus, for the two spacelike separated
particles locality is a necessary and sufficient condition for the states
having the form specified in equation (\ref{sep}). Consequently, the state
(\ref{sep}) is local by definition. On the contrary, any state that cannot be
written as equation (\ref{sep}) is nonlocal.

Importantly, since the above reasoning does not require any specification of
the particles, it is general enough that it is valid for any biparticle
systems. Furthermore, its generalization to multiparticle systems is
straightforward, and the resulting states satisfying locality are still given
by a convex sum of direct products of the local density operators, similarly
to equation (\ref{sep}).

Surprisingly, the form of $\rho_{AB}$ in equation (\ref{sep}) is just the
\textquotedblleft mathematical\textquotedblright\ definition of separable
(i.e., non-entangled) two-particle states, as is suggested by Werner
\cite{Werner} and now widely accepted. Meanwhile, $\rho_{AB}$ is local by
definition as it comes from the \textquotedblleft physical\textquotedblright%
\ criterion of locality, which clarifies the physical content, and justifies
Werner's definition, of separable states from another perspective.
Consequently, the two basic notions---entanglement of quantum states and
nonlocality of measured results---are equivalent for spacelike separated
systems; local measurements performed on entangled (separable) states give
nonlocal (local) results. This result is a further support of the viability of
the locality condition (\ref{local}). When the particles A and B are not
spacelike separated, the locality condition given in equation (\ref{local})
can be reasonably called Einstein's separability condition. In this broader
sense, locality is identical to Einstein's separability when the particles are
spacelike separated, and Einstein's separability is a necessary and sufficient
condition for the separability of states for any quantum systems.

This fact may open up an exciting perspective for understanding nonlocality
(or more generally, Einstein's inseparability) and entanglement as a unified
concept in quantum mechanics, i.e., quantum wholeness: Entanglement represents
the mathematical inseparability of quantum states, while nonlocality
physically manifests itself in the correlations of certain measurement
results. Thus, one has to accept the existence of a quantum weirdness in
nature: Quantum entanglement induces exotic influences for a composite system
even when the constituent parts are spacelike separated. Here one is
confronted with a situation where an entangled quantum system must be regarded
as a holistic entity; any attempt to describe the entangled system locally
must fail for certain quantum predictions.

As we proved, locality permits only the classically correlated (i.e.,
separable) states (\ref{sep}) in quantum mechanics. Or equivalently, whatever
the set of common causes is, the locality assumption (\ref{local}) cannot be
fulfilled by any entangled state. However, entanglement is ubiquitous in
quantum mechanics as well as in practical quantum information processing as an
essential resource. Thus, locality is in conflict with quantum mechanics,
namely, quantum mechanics is definitely a nonlocal theory. We believe that we
have for the first time proved in simple terms the intrinsic nonlocal feature
of quantum mechanics in a clear-cut way, without resorting to the realism
assumption \cite{Bell,Bell-book} or other counterfactual reasonings
\cite{Stapp,Vaidman}. Actually, the realism assumption used in deriving Bell's
inequalities is redundant and even detrimental for the purpose of uncovering
quantum nonlocality \cite{Chen-test}.

One might wonder whether the apparent nonlocality of quantum mechanics could
be used for superluminal signaling. If this were the case, then quantum
mechanics would violate the relativistic causality which forbids any
superluminal causal action. Fortunately, quantum mechanics in its current form
is still in a \textquotedblleft peaceful co-existence\textquotedblright\ with
relativity \cite{Shimony} in the sense that nonlocality does not lead to
superluminal information transmitting. To see this, recall that the locality
assumption (\ref{local}), according to Shimony \cite{Shimony}, consists of two
independent factors: (i) the outcome independence and (ii) the parameter
independence, which demand that any measurement outcome of particle A should
be independent, respectively, on the outcomes and on the experimental settings
of the spacelike separated particle B. Explicit calculation shows that quantum
mechanics violates the outcome independence, which nevertheless cannot be used
for superluminal information transmitting \cite{Peres-book,Shimony}. However,
violation of the parameter independence may imply superluminal communication
\cite{assume,Shimony}. To prove the parameter independence (see, e.g.,
\cite{Shimony,Laloe}) in the two-particle case (the multiparticle
generalization is straightforward), it is sufficient to prove \cite{assume}
$P(a_{i}\left\vert b)\right.  \equiv\sum_{j}P(a_{i},b_{j})=P(a_{i})$ for any
chosen setting $b$.\ Actually, this is always true as\ $\sum_{j}P(a_{i}%
,b_{j})=Tr(\rho_{AB}\hat{P}_{Ai}\sum_{j}\hat{P}_{Bj})=Tr(\rho_{A}\hat{P}%
_{Ai})=P(a_{i})$. Here we have used the simple fact that $\sum_{j}\hat{P}%
_{Bj}=\hat{I}_{B}$ ($\hat{I}_{B}$ is the unit operator for particles B);
$\rho_{A}=Tr_{B}(\rho_{AB})$ is the reduced density operator for particle A.
Thus, quantum mechanics respects the parameter independence, implying that
quantum nonlocality cannot result in superluminal signaling.

Based on the equivalence between nonlocality/inseparability of measurements
and entanglement of states, quantum nonlocality has acquired the same solid
basis as quantum entanglement. The two equivalent notions are two distinct
aspects of the same quantum wholeness. This fact, being interesting in its own
right on fundamental issues of quantum mechanics, might be important in
quantum information science, where manipulating entanglement is a vital task
for processing information. This practical impulse has greatly enriched our
current knowledge on entanglement. We anticipate that future works on
entanglement and nonlocality may be mutually promoted to deepen our
understanding of the weird quantum wholeness.

We thank Jian-Wei Pan and Lu-Ming Duan for useful discussions. This work was
supported by the National Natural Science Foundation of China, the Chinese
Academy of Sciences and the National Fundamental Research Program. N.-L.L. is
also supported by the Scientific Research Foundation for the Returned Overseas
Chinese Scholars by the State Education Ministry of China and the USTC
Returned Overseas Scholars Foundation.


\begin{thebibliography}{99}
\bibitem {note}One can prove the following fact: If $\left\langle
\phi\right\vert \left\langle \psi\right\vert \hat{O}\left\vert \phi
\right\rangle \left\vert \psi\right\rangle =0$\ for any product state
$\left\vert \phi\right\rangle \left\vert \psi\right\rangle $ for a
two-particle system (generalization to multiparticle cases is
straightforward), then the two-particle operator $\hat{O}\equiv0$. To this
end, let $\left\vert i\right\rangle $ be any state of particle A, and
$\left\vert k\right\rangle $ and $\left\vert l\right\rangle $ any two
orthonormal states of particle B. Then for the product states $\left\vert
\varphi_{1}\right\rangle =\left\vert i\right\rangle \otimes\frac{1}{\sqrt{2}%
}\left(  \left\vert k\right\rangle +\left\vert l\right\rangle \right)  $ and
$\left\vert \varphi_{2}\right\rangle =\left\vert i\right\rangle \otimes
\frac{1}{\sqrt{2}}\left(  \left\vert k\right\rangle +i\left\vert
l\right\rangle \right)  $, one necessarily has $\left\langle \varphi
_{1}\right\vert \hat{O}\left\vert \varphi_{1}\right\rangle =\frac{1}{2}\left(
\left\langle ik\right\vert \hat{O}\left\vert ik\right\rangle +\left\langle
ik\right\vert \hat{O}\left\vert il\right\rangle \right.  \left.  +\left\langle
il\right\vert \hat{O}\left\vert ik\right\rangle +\left\langle il\right\vert
\hat{O}\left\vert il\right\rangle \right)  $\ and $\left\langle \varphi
_{2}\right\vert \hat{O}\left\vert \varphi_{2}\right\rangle =\frac{1}{2}\left(
\left\langle ik\right\vert \hat{O}\left\vert ik\right\rangle +i\left\langle
ik\right\vert \hat{O}\left\vert il\right\rangle \right.  \left.
-i\left\langle il\right\vert \hat{O}\left\vert ik\right\rangle +\left\langle
il\right\vert \hat{O}\left\vert il\right\rangle \right)  $. Since
$\left\langle \varphi_{1}\right\vert \hat{O}\left\vert \varphi_{1}%
\right\rangle =\left\langle \varphi_{2}\right\vert \hat{O}\left\vert
\varphi_{2}\right\rangle =\left\langle ik\right\vert \hat{O}\left\vert
ik\right\rangle =\left\langle il\right\vert \hat{O}\left\vert il\right\rangle
=0$, one obtains from the above equations that $\left\langle ik\right\vert
\hat{O}\left\vert il\right\rangle =0$. Particularly, choosing $\left\vert
i\right\rangle $ to be the eigenvectors of the Hermitian operator
$\left\langle k\right\vert \hat{O}\left\vert l\right\rangle +\left\langle
k\right\vert \hat{O}\left\vert l\right\rangle ^{\dagger}$ for particle A and
using $\left\langle ik\right\vert \hat{O}\left\vert il\right\rangle =0$, it is
easy to see that $\left\langle k\right\vert \hat{O}\left\vert l\right\rangle
+\left\langle k\right\vert \hat{O}\left\vert l\right\rangle ^{\dagger}=0$.
Similarly, one can prove $\left\langle k\right\vert \hat{O}\left\vert
l\right\rangle -\left\langle k\right\vert \hat{O}\left\vert l\right\rangle
^{\dagger}=0$. Therefore, $\left\langle k\right\vert \hat{O}\left\vert
l\right\rangle =0$, which means that $\hat{O}\equiv0$. In particular, for
$\hat{O}=\rho_{AB}-\sum_{\mu}\lambda_{\mu}\rho_{A\mu}\rho_{B\mu}$, $\left\vert
\phi\right\rangle =\left\vert a_{i}\right\rangle $ and $\left\vert
\psi\right\rangle =\left\vert b_{j}\right\rangle $, equation (\ref{roab})\ and
the result proved in this note then give equation (\ref{sep}).
\end{thebibliography}
\end{document}